\documentclass[prb,twocolumn,preprintnumbers,amsmath,amssymb,superscri
ptaddress]{revtex4}
\usepackage{graphicx}
\usepackage{latexsym}
\usepackage{amsmath}
\usepackage{graphics}
\usepackage{amssymb}
\usepackage{layout}
\usepackage{verbatim}
\usepackage{amsfonts,epsfig}
\newcommand{\ket}[1]{\left|#1\right>}
\newcommand{\bra}[1]{\left< #1 \right|}
\newcommand{\beq}{\begin{equation}}
\newcommand{\eeq}{\end{equation}}

\newcommand{\mean}[1]{\langle{#1}\rangle{}}

\begin{document}

\title{Spin echo decay at low magnetic fields in a nuclear spin bath}
\author{{\L}ukasz Cywi{\'n}ski}
\email{lcyw@ifpan.edu.pl}
\affiliation{Institute of Physics, Polish Academy of Sciences,
al.~Lotnik{\'o}w 32/46, PL 02-668 Warszawa, Poland}
\affiliation{Condensed Matter Theory Center, Department of Physics,
University of Maryland, College Park, MD 20742-4111, USA}
\author{V.~V. Dobrovitski}
\affiliation{Ames Laboratory, Iowa State University, Ames, Iowa 50011,
USA}
\author{S. Das Sarma}
\affiliation{Condensed Matter Theory Center, Department of Physics,
University of Maryland, College Park, MD 20742-4111, USA}
\date{\today }

\begin{abstract}
We investigate theoretically the spin echo signal of an electron
localized in a quantum dot and interacting with a bath of nuclear
spins. We consider the regime of very low magnetic fields
(corresponding to fields as low as a militesla in realistic GaAs and
InGaAs dots). We use both the exact numerical simulations and the
analytical theory employing the effective pure dephasing Hamiltonian.
The comparison shows that the latter approach describes very well the
spin echo decay at magnetic fields larger than the typical Overhauser
field, and that the timescale at which this theory works is larger
than previously expected. The numerical simulations are also done for
very low values of electron spin splitting at which the effective Hamiltonian based 
theory fails quantitatively. Interestingly, the qualitative difference in the spin
echo decay between the cases of a homonuclear and a heteronuclear bath
(i.e.~bath containing  nuclear  isotopes having different Zeeman
energies), predicted previously using the effective Hamiltonian
approach, is still visible at very low fields outside the regime of
applicability of the analytical theory. We have found that the spin
echo signal for a homonuclear bath oscillates with a frequency
corresponding to the Zeeman splitting of the single nuclear isotope
present in the bath. The physics behind this feature is similar to that of the
electron spin echo envelope modulation (ESEEM). While purely isotropic
hyperfine interactions are present in our system, the tilting of the
electron precession axis at low fields may explain this result.
\end{abstract}

\maketitle

\section{Introduction}

The problem of the dynamics of an electron spin coupled by a
hyperfine (hf) interaction to a bath of nuclear spins has
been a focus of much theoretical attention, since the
interaction with the nuclear bath is the most limiting
decoherence mechanism in spin qubits based on quantum dots
made of III-V materials.\cite{Hanson_RMP07,Coish_pssb09}
While at high magnetic fields ($B \! \gg \! 0.1$ T in large
GaAs dots when spin echo is considered)
the dipolar interactions between the nuclear spins are the
main source of dynamics leading to decoherence seen in spin echo
(SE)
experiment,
\cite{deSousa_PRB03,Witzel_PRB05,Witzel_PRB06,Yao_PRB06,Witzel_PRB08}
at lower fields the electron spin
dephases and relaxes due to hf interaction alone. This
process has been studied theoretically for free evolution or
SE both by analytical
methods
\cite{Khaetskii_PRB03,Coish_PRB04,Yao_PRB06,Zhang_PRB06,Liu_NJP07,Saikin_PRB07,Deng_PRB06,Deng_PRB08,Coish_PRB08,Cywinski_PRL09,Cywinski_PRB09,Coish_PRB10}
and by exact numerical
simulations.\cite{Schliemann_JPC03,Dobrovitski_PRE03,Shenvi_scaling_PRB05,Zhang_PRB06,Zhang_JPC07,Zhang_Viola_PRB08}
Here we focus on the theoretical description of SE decay due
to hyperfine interaction
alone,\cite{Shenvi_scaling_PRB05,Yao_PRB06,Liu_NJP07,Saikin_PRB07,Cywinski_PRL09,Cywinski_PRB09} since the SE experiment
is currently the most developed measurement of coherence
decay in electrically controlled gated GaAs quantum
dots,\cite{Petta_Science05,Koppens_PRL08,Bluhm_arXiv10} and there
has been a recent progress in performing SE measurements on
optically controlled electron spins in quantum
dots\cite{Greilich_NP09,Press_NP10} and spins of electrons bound to
donors.\cite{Clark_PRL09}

Most of  the analytical approaches were concentrated on the
``perturbative''
regime\cite{Coish_PRB04,Deng_PRB06,Deng_PRB08,Coish_PRB08,Coish_PRB10}
of magnetic fields in which the electron Zeeman energy $\Omega$ and
the total hf energy $\mathcal{A} \! \equiv \! \sum_{i}A_{i}$ (where
the sum is over all the nuclei and $A_{i}$ are the individual hf
couplings) fulfill $\mathcal{A}/\Omega \! \ll \! 1$. Only recently it
has been proposed\cite{Cywinski_PRL09,Cywinski_PRB09} that an
analytical calculation can be  well-controlled under  a much weaker
condition of $\delta \equiv \mathcal{A}/\Omega\sqrt{N} \! \ll \! 1$
($N$ being the number of nuclei interacting with the central spin),
allowing the calculation of decoherence at much lower $B$ fields
(which only have to be larger than the typical Overhauser field due to
the nuclei, which is on the order of a few mT in large GaAs dots). In
this theory, as in earlier closely related
studies,\cite{Yao_PRB06,Liu_NJP07,Saikin_PRB07} an effective pure
dephasing Hamiltonian describing hyperfine-mediated interactions
between the nuclei is used. For a pure dephasing problem one can
formulate a diagrammatic expansion technique for spin decoherence time
evolution, and a class of diagrams of leading order in $1/N$ expansion
can be resummed,\cite{Cywinski_PRB09} leading to predictions for
narrowed state\cite{Klauser_PRB06,Greilich_Science06}
Free Induction Decay (FID), SE decay, and also decoherence under any
other dynamical
decoupling\cite{Viola_JMO04,Yao_PRL07,Witzel_PRL07,Witzel_CDD_PRB07,Zhang_Viola_PRB08} sequence of ideal $\pi$ pulses driving the qubit.

In this article we present a comparison between the SE decay
calculated using the ring diagram theory (RDT) of
Refs.~\onlinecite{Cywinski_PRL09,Cywinski_PRB09} and exact simulations
of a system with $N\! = \! 20$ nuclei.
In this way we set out to clarify the limits of quantitative
applicability of the RDT, i.e.~the ranges of electron spin splitting
and time-scales on which the analytical theory based on effective
Hamiltonian accurately describes the SE decay. We also investigate
numerically the regime of very low spin splittings, at which the RDT
is bound to fail, and we analyze simplified approaches which can be used
to model (at least on a certain timescale) the SE signal in this
regime.

The exact results confirm that the RDT is quantitatively
accurate as long as $\delta \! \ll \! 1$, and
that the presence of multiple nuclear isotopes (having
different Zeeman splittings) is crucial for qualitatively
correct description of the SE decay at these low fields. While with
$N\! = \! 20$ spins used in the exact calculation it is impossible to
unequivocally discern whether it is $\delta \! \ll \! 1$ or
$\mathcal{A}/\Omega \! \ll \! 1$ which is controlling the quantitative
agreement, our simulations strongly suggest that the
qualitative (or even semi-quantitative) predictions of the RDT still
hold even at lower $B$ fields. We observe that RDT correctly predicts
many qualitative features of the dephasing process even for $\delta
\sim 1$, i.e.\ the RDT results are valid (at least qualitatively, and
at least at certain timescale $\tau_R$) beyond the regime $\mathcal{A}
/\Omega\ll 1$.
The striking qualitative difference in SE decay
between a heteronuclear and a homonuclear system is clearly
visible for $\delta \! \approx \! 1$ (when $\mathcal{A}/\Omega \! > \!
1$), and the timescale at which the majority of the decay (say, by
half of the initial amplitude) occurs is reproduced by the RDT. This
leads us to the conclusion that the smallness of $\mathcal{A}/\Omega$
is not necessary for correct analytical description of the spin echo
signal.

Another interesting feature of the low-$B$ SE signal
in a homonuclear system is the
oscillation of this signal with time at a frequency given by nuclear
Zeeman
splitting. This feature is clearly visible in our results, and it
should not be confused with the oscillations appearing in
a heteronuclear situation, when the frequencies are given
by the \emph{differences} of nuclear Zeeman energies of
different isotopes.\cite{Cywinski_PRL09,Cywinski_PRB09,Bluhm_arXiv10}

The paper is organized in the following way. In Sec.~\ref{sec:theory}
we give a brief outline of the analytical RDT theory and its
limitations, and we describe the system of 20 nuclei to which an exact
numerical method of evaluation of spin dynamics is applied. In
Sec.~\ref{sec:results} we present the results of the calculations, and
in Sec.~\ref{sec:discussion} we discuss them and compare them with
simplified ``box wavefunction'' model (in which all the hf couplings
are the same). In Appendix \ref{app:choice} we provide an extended discussion of the influence of choice of the hf couplings on the results of the calculations in a system with $N\! = \! 20$ nuclei.

\section{Theoretical approach}  \label{sec:theory}
The Hamiltonian is given by
\beq
\hat{H} = \Omega \hat{S}^{z} + \sum_{i} \omega_{\alpha[i]}\hat{J}^{z}
_{i} + \sum_{i} A_{i} \mathbf{S}\cdot \mathbf{J}_{i} \,\, ,
\label{eq:H}
\eeq
where $\Omega$ is the electron spin Zeeman splitting, $
\omega_{\alpha[i]}$ is the Zeeman energy of the $i$-th nuclear spin
which belongs to nuclear species $\alpha$, and the last term is the hf
interaction. We employ $J\!=\! 1/2$ nuclear spins in the paper.

\subsection{Overview of the effective-Hamiltonian-based theory of ring
diagram resummation}
When $\Omega$ is large enough one can perform a canonical
transformation on the full hf Hamiltonian given in Eq.~(\ref{eq:H}),
which removes the electron-nuclear spin flip terms ($\hat{S}^{\pm}
\hat{J}^{\mp}_{i}$) in favour of a hierarchy of intra-bath hf-mediated
interactions involving two or more nuclear
spins.\cite{Shenvi_scaling_PRB05,Yao_PRB06,Coish_PRB08,Cywinski_PRL09,
Cywinski_PRB09} It has been argued\cite{Cywinski_PRL09,Cywinski_PRB09}
that as long as $\delta \! \ll \! 1$ and when considering the
evolution up to a certain time-scale $\tau_{R}$, one needs to take
into account only the lowest-order term in the the expansion of the
effective hamiltonian $\tilde{H}$:
\begin{eqnarray}
\tilde{H}^{(2)} \!\! & =  & \!\! -\sum_{i} \frac{A^{2}_{i}}{4\Omega}
\hat{J}^{z}_{i}  + \hat{S}^{z}\sum_{i} \frac{A^{2}_{i}}{4\Omega}  +
\hat{S}^{z}\sum_{i\neq j} \frac{A_{i} A_{j}}{2 \Omega} \hat{J}^{+}_{i}
\hat{J}^{-}_{j}  \,\, , \label{eq:H2}
\end{eqnarray}
where the first two terms are renormalizations of the nuclear and
electron Zeeman energies, respectively, and the third term is the  hf-mediated interaction.

With such an effective \emph{pure dephasing} Hamiltonian (where the
only electron spin operator present is $\hat{S}^{z}$) the diagonal
elements of  the reduced density matrix of the central spin (giving
the average spin in the $z$ direction) remain constant in time, and
the evolution of the off-diagonal element $\rho_{+-}(t)$ can be mapped
onto evaluation of the bath-averaged contour-ordered
exponent.\cite{Grishin_PRB05,Saikin_PRB07,Yang_CCE_PRB08,Lutchyn_PRB08,Cywinski_PRL09,Cywinski_PRB09} This formulation of the problem allows
one to employ some of the tools of the diagrammatic perturbation
theory, most importantly the linked cluster theorem, using  which one
can write $\rho_{+-}$ as an exponent of the sum of all the connected
diagrams contributing to the original expansion.

The  third term in Eq.~(\ref{eq:H2}) is a long-range interaction
coupling \emph{all} the $N$ nuclei within the bulk of the electron's
wavefunction with comparable strength. For such an interaction all the
terms in the perturbative expansion of $\rho_{+-}(t)$ can be
classified by their $1/N$ dependence. The leading-order terms can then
be resummed and the time-dependence of the SE signal can be easily
calculated.\cite{Cywinski_PRL09,Cywinski_PRB09} Let us stress that
both the exponential resummation of the original expression for
$\rho_{+-}(t)$ and further resummation of a class of linked diagrams
are technically feasible because we use an approximate effective pure
dephasing Hamiltonian.

The decoherence due to the bath dynamics caused by the dipolar
interactions between the nuclear spins (spectral diffusion) can can be
calculated in a similar way, but only the second order (in dipolar
coupling) diagram needs to be retained in the cluster expansion in
order to get a controlled description of SE decay,
\cite{Witzel_PRB06,Yao_PRB06,Saikin_PRB07} and a very good agreement
with spin echo measurements of spins of phosphorus donors in Si was
obtained.\cite{Witzel_PRB06,Tyryshkin_JPC06,Witzel_AHF_PRB07}
Let us also mention that the theoretical
prediction\cite{Witzel_PRB05,Witzel_PRB06,Yao_PRB06} of the coherence
decay timescale of $\gtrsim \! 10$ $\mu$s due to the spectral
diffusion at high $B$ in GaAs quantum dots was recently
confirmed.\cite{Bluhm_arXiv10}
According to the RDT, in GaAs dots at sub-Tesla magnetic field the hf-
mediated interactions give the SE decay which is an order of magnitude
faster than the decay due to dipolar
interactions,\cite{Cywinski_PRL09,Cywinski_PRB09} and since we are
interested in correct description of coherence dynamics on timescale
comparable to its characteristic decay time, we neglect the dipolar
dynamics altogether in the low-$B$ regime in which our focus is on
relatively short times.

For the relevant here case of spin echo we define the decoherence
function $W(t) \! \equiv  \! \rho_{+-}(t)$ (assuming $\rho_{+-}(0) \! =\! 1$), with which
the expectation values of the transverse components of the central
spin are given by $\mean{\hat{S}^{x}(t)} \! =\! \frac{1}{2} \text{Re}
W(t)$ and $\mean{\hat{S}^{y}(t)} \! =\! -\frac{1}{2} \text{Im} W(t)$.

The RDT calculation of decoherence due to two-spin hf-mediated
interactions simply involves diagonalization of $N \! \times \! N$
matrix at each time step. Such a $T$-matrix can be easily calculated,
as described in details in Ref.~\onlinecite{Cywinski_PRB09}, and the
decoherence function is given then by the following formula involving
the eigenvalues $\lambda_{l}(t)$ of the $T$-matrix:
\beq
W(t) = \prod_{l}^{N} \frac{1}{\sqrt{1+\lambda^{2}_{l}(t)}} \,\, .
\eeq
Let us note that the pair correlation approximation (PCA) from
Refs.~\onlinecite{Yao_PRB06,Yao_PRL07,Liu_NJP07} can be extended to
the hetero-nuclear case considered here, and it leads to approximating
the decoherence function by $W_{\text{PCA}} \! = \! \exp ( -\frac{1}
{2}\sum_{l} \lambda^{2}_{l} )$. This corresponds to taking only the
lowest order term in the linked cluster
expansion\cite{Saikin_PRB07,Cywinski_PRB09} of $W(t)$, while in the
RDT we resum all the terms in this expansion which are of the leading
order in $1/N$ at every order in the spin-spin coupling.

The key feature of the spin echo sequence in the high-field  ($\delta
\! \ll \! 1$) regime is that the dephasing of the central spin due to
$\hat{S}^{z}$-conditioned  interaction from Eq.~(\ref{eq:H2}) with a
single nuclear species (i.e.~all the nuclei having the same Zeeman
splitting) is practically completely undone by the pulse
sequence.\cite{Shenvi_scaling_PRB05,Yao_PRB06} However, at magnetic
field lower that $B_{c} \! \approx \! \sqrt{\Omega/\Delta\omega}
\mathcal{A}/\sqrt{N}|g_{\text{eff}}|\mu_{\text{B}}$ (where
$\Delta\omega$ is the typical difference of Zeeman splittings between
distinct nuclear species and $g_{\text{eff}}$ is the effective g-
factor of the electron) the hf-mediated processes between nuclei of
different species were predicted to completely dominate the SE
decay.\cite{Cywinski_PRL09,Cywinski_PRB09} This prediction has
recently been confirmed experimentally in a spin echo experiment in a
double dot singlet-triplet qubit,\cite{Bluhm_arXiv10} in which the
$B$-dependence of the SE decay and the characteristic oscillations of
the SE signal with frequencies $\propto \omega_{\alpha\beta} \equiv
\omega_{\alpha}-\omega_{\beta}$ have been seen. Let us note that with
the parameters used in the calculations below the $B$ field regime in
which these heteronuclear process dominate the decay corresponds to
$\Omega \! < \! \Omega_{c} \! \approx \! 100$ in the units defined
below.

\subsection{Approximations inherent in the ring diagram calculation}
There are higher-order terms in the expansion of the effective
Hamiltonian $\tilde{H}$, which correspond to virtual processes
involving more than two $S$-$J$ spin flips.\cite{Cywinski_PRB09} These
terms also involve increasing numbers of nuclear spins, i.e.~there are
$n$-spin interactions in $\tilde{H}^{(n)}$, while the RDT can only
deal with two-spin interactions. The exact derivation of these terms
quickly becomes very complicated, but we do not expect such an
expansion to be useful: when the first higher-order interaction in
$\tilde{H}$ becomes important, all the higher order terms become
equally relevant. However, in order to illustrate the breakdown of the
RDT we will use in the calculations the two-spin $\hat{S}^{z}$-
independent interaction which appears in $\tilde{H}^{(3)}$, in which
the coupling constants are proportional to $\mathcal{A}^{3}/N^{2}
\Omega^{2}$.

A more important approximation underlying RDT is the fact that the
effective-Hamiltonian approach neglects the
transformation of states associated with the canonical transformation
(i.e.~only the Hamiltonian is transformed). As a result, no decay of
$S^{z}$ component of the central spin can be obtained. This also has
consequences for the decay of the transverse spin components at low
$B$ fields, where the effective-Hamiltonian based theory does not
reproduce the
fast ``visibility loss'' process.\cite{Coish_PRB04,Shenvi_scaling_PRB05,Yao_PRB06} The latter process, in which the
transverse spin component decays by $\sim \delta^{2}$ on a time-scale
of
$\sqrt{N}/\mathcal{A}$ can be captured by an exact simulation or by a
theory
in which at least some effects of state transformation are retained,
see e.g.~Ref.~\onlinecite{Yao_PRB06}. A more detailed calculation of
this process, which is associated with decay of the $S^{z}$ component
of the central spin, is given in Ref.~\onlinecite{Coish_PRB04}, where
the Generalized Master Equation method with the full hf Hamiltonian
was used.

The transformation of states corresponds to an entanglement of
electron spin with the nuclei (each possible initial bath state gets
entangled on a short time-scale with the central spin state).
Semiclassically it can be envisioned as a slight tilting of the
electron quantization axis from the $z$ direction, which
happens
due to hf interaction with the collective spin of the nuclei. The
latter is semiclassicaly reproduced as a random vector, and the
tilting of the central spin's
$z$-axis is conditioned upon the size and the direction of this
vector.

The visibility loss is only one example of effects which are absent in
an
effective-Hamiltonian approach. There are also other
features, which require the treatment based on the full hf
Hamiltonian. The analytical approaches of this kind have
only been applied to the case of FID decay in a narrowed nuclear
state.\cite{Coish_PRB04,Deng_PRB06,Deng_PRB08,Coish_PRB10} At times
much longer than $N/\mathcal{A}$ an  asymptotic $1/t^2$ coherence
decay was obtained,\cite{Deng_PRB06,Deng_PRB08,Coish_PRB10} which is
not reproduced by the RDT
calculation.\cite{Cywinski_PRL09,Cywinski_PRB09} However at high
fields ($\mathcal{A}/\Omega \! < \! 1$) most of the narrowed FID decay
is described by an exponential $e^{-t/T_{2}}$ obtained both by the RDT
(and PCA at very high fields\cite{Liu_NJP07}) and by the Generalized
Master Equation (GME) approach using the full Hamiltonian in
Ref.~\onlinecite{Coish_PRB10}, and using the effective Hamiltonian in
Ref.~\onlinecite{Coish_PRB08}.

It should be stressed that in the full-Hamiltonian theories it was
argued that the convergence of the calculation is guaranteed only when
$\mathcal{A}/\Omega$ is small, which is a much more restrictive
requirement than $\delta \ll 1$ required in RDT theory. However,
neglecting the higher-order multi-spin interactions in the effective
Hamiltonian was argued\cite{Cywinski_PRB09} to be a good approximation
for $\delta \! \ll \! 1$  \emph{ only up to a certain time-scale}
$\tau_{R}$, so that even if the effects of the state transformation
were unimportant, the temporal regime of applicability of RDT is
certainly limited. On the other hand, the GME theories seem to give a
well-controlled result for \emph{all times}, but in a more restricted
regime of magnetic fields. Let us also mention that $\tau_{R}$ might
be different for narrowed FID and SE, so that the conclusions on the
relation of our results on SE presented here to the results obtained
for narrowed FID with other theories should be drawn carefully.

An experimentally relevant question is whether $\tau_{R}$ is larger
than the timescale of significant coherence decay. While the estimate
of $\tau_R \! \sim \! N/\mathcal{A}$ given previously is currently
enough for making contact with recent
experiments\cite{Petta_Science05,Koppens_PRL08,Bluhm_arXiv10} on spin
echo in GaAs where it corresponds to about $10$ $\mu$s, establishing a
more precise bound is of large current theoretical and possibly future
experimental interest.

\subsection{Numerical treatment of the full hyperfine Hamiltonian}
In order to establish more firmly the regime of applicability of the
RDT one should compare its predictions with exact calculations
starting from the full hf Hamiltonian from Eq.~(\ref{eq:H}). In this
way it is also possible to investigate the regime of very low $B$
fields (corresponding to $\delta \! > \! 1$), in which no
comprehensive analytical approach has been successful.
In the exact numerical simulation the time-dependent Schr{\"o}dringer
equation for the central spin and the bath is solved using the
Chebyshev polynomial expansion of the evolution
operator.\cite{Dobrovitski_PRE03,Zhang_JPC07}

Although the nuclear spins in GaAs and InAs have $J\!=\! 3/2$ or
$9/2$, we use $J\!=\! 1/2$ in the paper. The main expected effect of
such a simplification is slowing down of the decay (larger nuclear
spins are more efficient at decohering the central spin). It allows us
however to use the maximal number of nuclei $N \! = \! 20$ in an exact
simulation, diminishing thus the possibility that differences between
the numerics and RDT simply come from the failure of $1/N$ expansion
of diagrams describing various processes in the nuclear bath.

In a real quantum dot the number of nuclei interacting appreciably
with the electron is commonly defined as $N \equiv \mathcal{A}^{2}
/\sum_{i}A^{2}_{i}$. We identify this quantity with the finite $N$
used in the simulations. This leads to $\delta \! \equiv \!
\sqrt{\sum_{i}A^{2}_{i}}/\Omega$. Furthermore, we use energy units in
which $\sum_{i} A^{2}_{i} \! = \! 1$, so that $\delta$ is equal simply
to $1/\Omega$ and $\mathcal{A}=\sqrt{N}$ in these dimensionless units.
The time is then measured in units of $\sqrt{N}/\mathcal{A}$, which in
real GaAs dot with $N \! \approx \! 10^{6}$ corresponds to about 10 ns
(with $J=1/2$ the $T^{*}_{2}$ decay time of inhomogeneously broadened
FID\cite{Merkulov_PRB02,Zhang_PRB06} is equal to $\sqrt{8}$ in the
dimensionless units used in the figures).

Our calculations correspond to the following experimental procedure.
Initially the electron spin is assumed to be directed along the
$\hat{x}$ axis in the $\ket{+x}$ state. After evolution for time $t/2$
an instantaneous rotation by $\pi$ about the $\hat{x}$ axis is applied
to the central spin.
The expectation value of the $\hat{S}^{x}$ operator at the final time
$t$ is then the SE amplitude, plotted in all the figures as a function of the total pulse
sequence time $t$. The general formula is given by
\beq
\mean{\hat{S}^{x}(t)} =  \Big\langle \bra{+x} e^{i\hat{H}\tau}
\hat{\sigma}_{x}  e^{i\hat{H}\tau} \frac{ \hat{\sigma}_{x}}{2} e^{-
i\hat{H}\tau} \hat{\sigma}_{x} e^{-i\hat{H}\tau} \ket{+x} \Big\rangle
\label{eq:Sx}
\eeq
where $\mean{...} \! = \! \frac{1}{Z}\text{Tr}_{J}[...]$ is the
average over an unpolarized nuclear bath (with $Z\! = \! 2^{N}$), and
$\tau \! \equiv \! t/2$.

The calculations have been performed on a model system of $N\! = \!
20$ nuclear spins coupled to the central spin by hf Hamiltonian from
Eq.~(\ref{eq:H}). The hf couplings $A_{i}$ were drawn from a random
distribution, with constraint $\sum_{i=1}^{N} A^{2}_{i} \! = \! 1$. We
discuss the choice of such a distribution, and also the sensitivity of
the results to the specific choice of $A_{i}$, in Appendix
\ref{app:choice}. Here we remark that with $N\! = \! 20$ the SE
signals are very similar one to another for most the choices of the
set of $A_{i}$, and below we employ the set of $A_{i}$ corresponding
to a rather typical SE signal.
In the hetero-nuclear case we have divided the nuclei into $3$ species
with $10$, $6$, and $4$ members in order to mimic the ratios of
concentrations of various isotopes in GaAs ($^{75}$As, $^{69}$Ga, and
$^{71}$ Ga, respectively). The Zeeman splittings $\omega_{\alpha}$ of
these $3$ species were fixed at $0.02526$, $0.0354$, and $0.045$,
again mimicking the ratios of nuclear Zeeman energies in GaAs.
Note that the $\omega_{\alpha}$ are kept fixed while the electron
Zeeman splitting $\Omega$ is varied. Although this does not correspond
to realistic situation when all the Zeeman energies are proportional
to the $B$ field, it allows us to more clearly separate the effects
that the electron and nuclear Zeeman energies have on the time
dependence of the SE signal. Let us note that while the ratio
$\Omega/\omega \! \sim \! 10^{3}$ in GaAs, here we are using
$\Omega/\omega \! \sim \! 10-10^{2}$ for the range of $\Omega$
considered below.
Another difference in comparison with the realistic GaAs dot is that
the differences of hf couplings $A_{ij} \! \equiv \! A_{i}-A_{j}$
(i.e.~the differences of Knight shifts of different nuclei) are of the
same order of magnitude or larger than the nuclear Zeeman energy differences
$\omega_{ij}$, while in GaAs dot with $N \! > \! 10^{5}$ we have
$A_{ij} \! \ll \omega_{ij}$ in the whole range of $B$ fields for which
$\delta \! \ll \! 1$. Thus, unless we put all the $A_{i}$ equal one to
another, we do not expect to see the $\omega_{ij}$-related oscillation
in the SE signal in the heteronuclear system (see
Ref.~\onlinecite{Cywinski_PRB09} for details on why the $A_{ij}\! \ll
\! \omega_{ij}$ condition is needed to obtain these oscillations). The
existence of this oscillation has been confirmed by recent
experiments,\cite{Bluhm_arXiv10} and here we focus on other aspects of
SE decay dynamics.

\section{Results for spin echo}  \label{sec:results}
First, let us briefly recount what we expect to see in exact numerics
at not-too-low magnetic fields ($\delta \! \ll \! 1$) based on
previous analytical work on spin echo. There should be a ``visibility
loss'' initial decay of SE signal of amplitude $\sim\! \delta^{2}$
occurring at timescale of $\sqrt{N}/\mathcal{A}$ (i.e. of order of
$O(N^0)$ in units
employed here), and we expect a qualitative difference in the
magnitude of SE decay between the homonuclear and heteronuclear bath.
The open questions are: what is the timescale $\tau_{R}$ on which the
RDT remain quantitatively accurate for $\delta \! \ll \! 1$, and what
happens at very low $B$ fields at which $\delta \! \sim \! 1$ or even
$\delta > 1$.

\begin{figure}[t]
\includegraphics[width=0.99\linewidth]{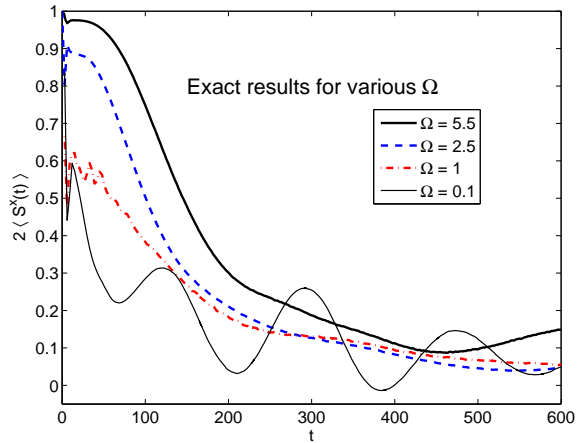}
\caption{(Color online) SE decay for $N\! = \! 20$ nuclei calculated
exactly for $\Omega \! = \! 0.1$, $1$, $2.5$, $5.5$. The hf couplings
$A_{i}$ are given as set 3 in Table \ref{tab:Ai}. The energy units are
such that $\sum_{i}A_{i}^2 \! = \! 1$. The corresponding time units
are such that $T_{2}^{*} \! =\! \sqrt{8}$ (for comparison, $T_{2}^{*}
\! \approx \! 10$ ns in GaAs dots). }  \label{fig:SE_exact}
\end{figure}

\begin{figure}[t]
\includegraphics[width=0.99\linewidth]{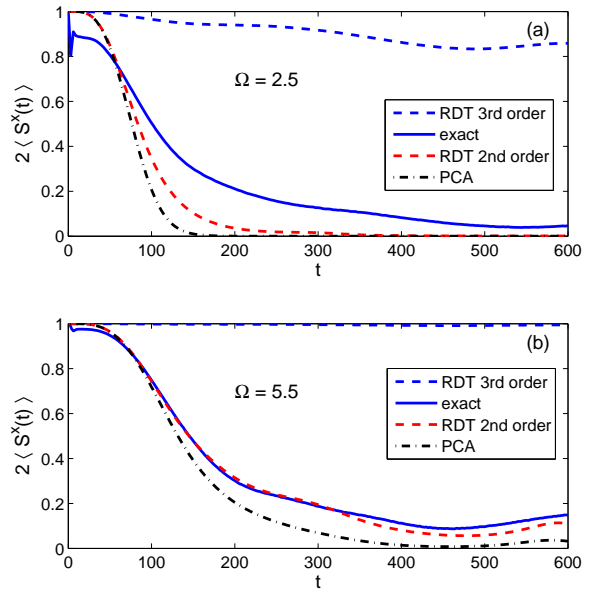}
\caption{(Color online) (a) Comparison between the exact (solid line)
result for $\Omega\! = \!2.5$, the result obtained using the RDT (red
dashed line for 2nd order $S^{z}$-conditioned interaction from
Eq.~(\ref{eq:H2}), blue dashed line for $S^{z}$-independent two-spin
interaction for the third order effective Hamiltonian), and the result
obtained using the Pair Correlation Approximation from
Ref.~\onlinecite{Yao_PRB06} (dot-dashed line). (b) same as (a) only
with $\Omega\! = \! 5.5$.} \label{fig:SE_hetero}
\end{figure}

In Figure \ref{fig:SE_exact} we are presenting the results of exact calculation for a hetero-
nuclear bath with electron Zeeman energies  $\Omega \! \in \! [0.1,
5.5]$ (corresponding to $\delta \! \in \! [0.18 ,10]$). As expected,
the exact calculation shows a very fast decay having $\delta^2$
magnitude for small $\delta$. In Figure \ref{fig:SE_hetero} we compare
the exact results for $\Omega \! = \! 2.5$ and $5.5$ with the RDT
calculations. These calculations were done using the $S^{z}$-
conditioned interaction from Eq.~(\ref{eq:H2}), and also with the
$S^{z}$-independent two-spin interaction appearing in the 3rd order of
the expansion of the effective Hamiltonian,\cite{Cywinski_PRB09} in
which the coupling constants are smaller by $\mathcal{A}/N\Omega$
factor. For $\Omega \! = \! 2.5$ one can see that this 3rd order
interaction is not completely irrelevant, signifying the importance of
higher-order corrections to the effective Hamiltonian approach. At
even lower values of $\Omega$ the RDT calculation fails to quantitatively describe
the decay : not only it does not capture the very fast initial drop,
but also at longer times it predicts a decay much faster than the one
given by the exact calculation (these results are not shown, but the
beginning of such disagreement between RDT and the exact signal can be
seen at $\Omega\! = \! 2.5$). 
The oscillatory character of the $\Omega
\! = \! 0.1$ signal will be discussed later in the paper.


On the other hand, at a slightly larger field $\Omega \! =\! 5.5$
(when $\delta \! = \! 0.18$), the RDT calculation using only the
lowest-order hf-mediated interaction is approximating very closely the
exact result, and the higher-order corrections are irrelevant. The
comparison between RDT and PCA in this case shows how the resummation
of all the ring diagrams extends the timescale on which the analytical
theory closely matches the exact calculation.
With $N\! = \! 20$ it is hard to say with full confidence whether it
is the smallness of $\mathcal{A}/\Omega$ or $\delta\! = \! \mathcal{A}
/\sqrt{N}\Omega$ which controls the agreement between the RDT
calculation and the exact result. However, it is clearly visible that
the onset of long-time agreement between the two calculations
correlates with the suppression of the short-time visibility loss,
which is known to be controlled by $\delta$. This strongly suggests
that it is indeed the smallness of $\delta$ that makes the RDT work.
Finally, these results are showing that  the timescale on which  the
RDT calculation of SE signal is quantitatively accurate visibly
exceeds a value of $\tau_{R} \! \approx \! N/\mathcal{A} \! \approx
\sqrt{20}$, which was conservatively estimated in
Ref.~\onlinecite{Cywinski_PRB09}.

\begin{figure}[t]
\includegraphics[width=0.99\linewidth]{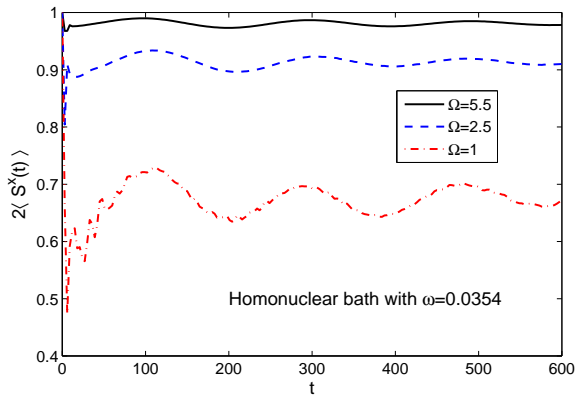}
\caption{(Color online) SE decay calculated exactly  for $N\! = \! 20$
with parameters as in Fig.~\ref{fig:SE_exact}, only with all nuclear
Zeeman energies set to $\omega\! = \! 0.0354$.
} \label{fig:SE_homo}
\end{figure}

\begin{figure}[t]
\includegraphics[width=0.99\linewidth]{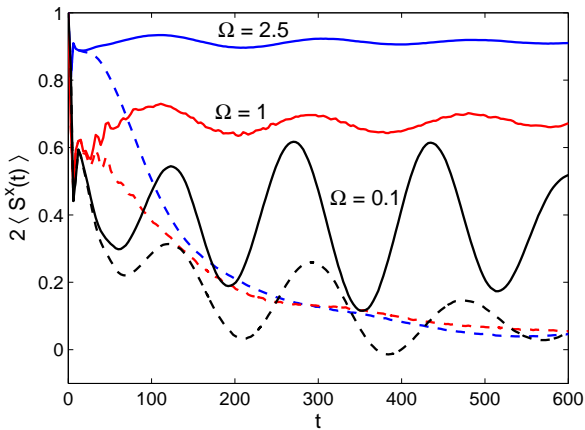}
\caption{(Color online) Comparison of the spin echo decay in a
heteronuclear bath (dashed lines, parameters the same as in
Fig.~\ref{fig:SE_exact}) and a homonuclear bath (solid lines,
parameters the same as in Fig.~\ref{fig:SE_homo}) for $\Omega \!= \!
2.5$, $1$, $0.1$ (with the initial decay smallest for the largest
$\Omega$).
} \label{fig:SE_homo_hetero}
\end{figure}

In order to illustrate the degree to which the low-field SE decay is
dominated by processes involving nuclei of different species, we have
performed the calculations for the homonuclear bath, in which we set
all the nuclear Zeeman energies equal to a common value $\omega$. In
Fig.~\ref{fig:SE_homo} we show the exact calculation for $\Omega=1$,
$2.5,$ and $5.5$. Now instead of the decay reaching $|W| \! < \! 0.1$
in Fig.~\ref{fig:SE_exact} we see only the initial visibility loss
followed by an oscillation with frequency given by the nuclear Zeeman
splitting $\omega$.

Within the RDT framework, the lowest-order term in $\tilde{H}$ which
contributes to the SE decay in the homonuclear bath is the  $S^{z}$-independent
two-spin interaction from $\tilde{H}^{(3)}$. The decay calculated by
RDT with this interaction for $\Omega \! = \! 2.5$ in the homonuclear
case is very similar to the result shown in Fig.~\ref{fig:SE_hetero}a
as blue dashed line, and it does not show any oscillation. The same
holds for $\Omega\!=\! 5.5$ case, where RDT predicts practically no
decay in the homonuclear case, while the exact calculation gives the
signal oscillating with the nuclear $\omega$ frequency around the
value of $1-\delta^{2}$.
Similar oscillations have been found before in numerical
simulations of the multi-pulse dynamical decoupling protocols for
an electron spin decohered by the nuclear spin
bath,\cite{Zhang_Viola_PRB08}
where full hyperfine interaction was considered.
This shows that this oscillation is a
feature following from the full $\mathbf{S}\cdot \mathbf{J}_{i} $
interaction between the electron and the nuclear spins, and as such
cannot be captured by the effective Hamiltonian theory. In this it
is similar to the fast initial oscillation with frequency
$\sim~\!\Omega$
accompanying the visibility loss. In the next section we discuss a
similarity (also noticed in Ref.~\onlinecite{Zhang_Viola_PRB08})
between this oscillation and the well-known electron spin echo
envelope modulation\cite{Mims_PRB72} (ESEEM).

In Fig.~\ref{fig:SE_homo_hetero} we show the comparison between the
exact SE decay in a hetero- and homonuclear bath at very low magnetic
fields ($\Omega \! \leq \! 2.5$), at which the RDT fails quantitatively. One can see
that the qualitative difference between the SE decay in the two cases
(homo- vs heteronuclear bath) is also clearly visible for $\Omega \! =
\! 1$, which is already completely outside the domain of applicability
of RDT. In this strongly non-perturbative regime the magnitude of the
initial decay ceases to be equal to $\delta^2$, and instead it is
smaller (with $|W| \! \sim \! 0.7$ after the initial decay). For the
homonuclear bath we then have a small-amplitude oscillation about this
value, while the decay in the heteronuclear case is practically
complete on the considered timescale. At longer times this decay is
very similar to the signal for $\Omega \! = \! 2.5$, which also shows
the robustness of qualitative predictions of RDT outside the regime of
its quantitative applicability: within RDT one obtains $B$-independent
SE decay below a certain value of magnetic field.\cite{Cywinski_PRB09}
At even lower values of $\Omega$ the situation changes: for $\Omega \!
= \! 0.1$ the homo- and heteronuclear case signals start to look
qualitatively similar. Although the physical picture of hf-mediated
interactions is not strictly applicable in this regime, one could
qualitatively describe this behavior by saying that the higher order
hf-mediated interactions (beyond the second order one, for which the
homo- and heteronuclear baths give qualitatively different SE decay)
become very strong, and the difference between the two cases
disappears. Thus, the oscillatory signal at $\Omega \! =\! 0.1$ has a
common origin in homo- and hetero-nuclear cases, and the discussion of
it is given in the next Section. Let us also note that the $\Omega\! =
0.1$ SE signal for the heteronuclear case does become negative at some
times.

\section{Discussion and comparison with a simple model}
\label{sec:discussion}
The presence of the oscillation with $\omega$ frequency in a
\emph{homonuclear} system can be derived using a simplified exactly
solvable model,\cite{Khaetskii_PRB03,Melikidze_PRB04,Zhang_PRB06} in
which all the hf couplings $A_{i}$ are put equal to the same value $A
\! = \! 1/\sqrt{N}$. The hf Hamiltonian is then given by $A\mathbf{S}
\! \cdot \! \mathbf{J}$ with $ \mathbf{J} \equiv \sum_{i} \mathbf{J}
_{i}$ being the operator of the total spin of the $N$ nuclei. The
calculation of the evolution of the system is most straightforward
using the basis of eigenstates of $\mathbf{J}^2$ and $\hat{J}^{z}$.
The only pairs of states coupled by the Hamiltonian are then
$\ket{\pm, j,m}$ and$\ket{\mp, j, m\pm 1}$, with the first quantum
number specifying the central spin state ($\pm$ for electron spin
up/down), and we have $\mathbf{J}^2\ket{j,m} \! = \! j(j+1) \ket{j,m}$
and $\hat{J}^{z}\ket{j,m} \! = \! m\ket{j,m}$.
Such a simplification of the dynamics of the system does not occur in
a heteronuclear case. There we have the full Hamiltonian given by
$\hat{H} = \Omega \hat{S}^{z} + \sum_{\alpha}\omega_{\alpha} \hat{J}
^{z}_{\alpha} + \sum_{\alpha}A_{\alpha}\mathbf{S}\cdot \mathbf{J}
_{\alpha}$, and even if we put $A_{\alpha}\!=\! A$ we must retain
distinct $\omega_{\alpha}$. If we then look at the dynamics starting
with the initial state, say, $\ket{+, \{ j_{\alpha},m_{\alpha} \} }$
(described by a set of $n_{J}$ pairs of quantum numbers
$j_{\alpha},m_{\alpha}$ for $\alpha \! = \! 1$...$n_{J}$), we see that
the Hamiltonian couples this state to a family of states with central
spin down and one $m_{\alpha}$ increased by $1$. Now, unlike in the
homonuclear case, these states can also couple to other states which
have the electron spin up and $m_{\beta}$ (with $\beta \! \neq \!
\alpha$) decreased by $1$. Thus, while in the homonuclear case we only
have to solve multiple two-state problems to obtain the system
dynamics, in the heteronuclear case we still have to consider Hilbert
spaces of dimensions as large as $2\prod_{\alpha} (N_{\alpha}+1)$
where $N_{\alpha}$ is the number of nuclei of $\alpha$ species.

\begin{figure}[t]
\includegraphics[width=0.99\linewidth]{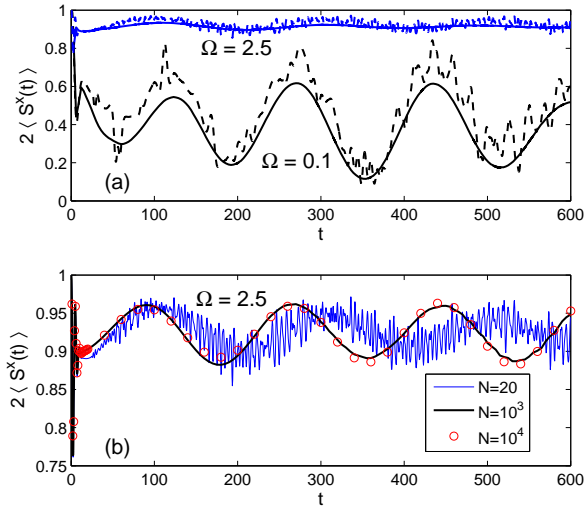}
\caption{(Color online) (a) The comparison between the exact results
(solid lines) for spin echo in a \emph{homonuclear} system (with the
nuclear Zeeman energy $\omega \! = \! 0.0354$) for electron spin
splitting $\Omega=2.5$, $0.1$ and the results obtained within the
model with all the hf couplings $A_{i}$ being the same (dashed lines).
(b) The comparison of such box-wavefunction calculations with
different numbers of nuclei $N$ for $\Omega\! = \! 2.5$. The fast
oscillation is an artifact of small $N$ and the box model, and it
disappears with increasing $N$.} \label{fig:box}
\end{figure}

The details of this calculation are given in Appendix \ref{app:box},
here we focus on the result shown in Fig.~\ref{fig:box}. In
Fig.~\ref{fig:box}a we present a comparison between the exact
calculation with nonuniform $A_{i}$ (solid lines) and the uniform
$A_{i}$ calculation (dashed lines). The slow oscillation with
frequency $\sim\! \omega$ and the amplitude of the signal is
reproduced by the ``box wavefunction'' calculation. The fast
oscillation (with frequency proportional to $\Omega$) visible in the
constant $A_{i}$ calculation is the small $N$ artifact related to
artificially regular structure of the Hamiltonian spectrum when all
$A_{i}$ are the same. In Fig.~\ref{fig:box}b we show the uniform
$A_{i}$ results for $N\! = \! 20$, $10^{3}$, and $10^{4}$. At larger
$N$ the fast oscillations disappear, and the shift of the oscillation
frequency, while visible, saturates quickly (the results for $N \! =\!
10^{3}$ and $10^{4}$ are practically the same). This allows us to
conclude that the presence of the $\omega$-oscillation is not a small
$N$ effect, and that such an oscillation should be present in the SE
signal of a central spin interacting with a large homonuclear bath.

For large $N$ the uniform $A_{i}$ model corresponds to the situation
in which a central spin $\mathbf{S}$ interacts with a large spin
$\mathbf{J}$. This leads us to a natural conjecture that the
$\omega$-oscillation follows from classical dynamics of the total
nuclear spin, i.e.~it can be recovered from classical equations of
motion for two spins coupled by isotropic Heisenberg coupling and
each experiencing a different Zeeman splitting. Since the typical
magnitude of the classical nuclear spin $\mathbf{J}$ is proportional
to $\sqrt{N}$, the electron spin precession due to hf coupling is
much faster than the nuclear spin precession due to interaction
with the electron spin.\cite{Merkulov_PRB02} This leads to
averaging out of the electron-induced nuclear precession, which
leaves only the nuclear spin precession due to the Zeeman term.
One is then looking at a problem of electron spin dynamics due
to the magnetic field and also due to the hf coupling with the
classical $\mathbf{J}$  vector precessing with frequency $\omega$
about the $z$ axis. The appearance of $\omega$ frequency in SE
signal is then natural. Our preliminary calculations involving
averaging of the classical dynamics of coupled $\mathbf{S}$ and
$\mathbf{J}$ spins over the ensemble of initial values of
$\mathbf{J}_{0}$ (without making the adiabatic approximation
used above) support this conjecture. This conjecture
is also in agreement with the physical picture suggested to explain
similar oscillations for multi-pulse dynamical decoupling
protocols.\cite{Zhang_Viola_PRB08}

Here we can comment on the $\Omega \! = \! 0.1$ result from
Figs.~\ref{fig:SE_exact} and \ref{fig:SE_homo_hetero}. At such a low
electron spin splitting the SE signal oscillations are present also
for the heteronuclear bath. These oscillations are controlled mostly
by the $3$ nuclear Zeeman frequencies (which we checked by varying
$\omega_{\alpha}$, not shown in the Figures), suggesting that the
physical picture described above, albeit with $3$ classical nuclear
spins, might be a starting point for the description of very low $B$
behavior of the spin echo.

For $\Omega \! \gg \! 1$ one can also see a close connection between
$\omega$-oscillation of the SE signal and the ESEEM, discussed
previously\cite{Witzel_AHF_PRB07} in the case of a central spin
interacting via \emph{anisotropic} hf interaction with the
nuclei,\cite{Saikin_PRB03} in which case the $\hat{J}^{x}\hat{S}^{z}$
coupling was a source of an oscillation of the SE signal. In our case,
if we neglect the ``visibility loss'' effect, the effective
Hamiltonian for $\Omega \! \gg \! 1$ is given by the sum of $\hat{H}
_{0}\! = \! \Omega\hat{S}^{z} + \omega\hat{J}^{z} + A\hat{S}^{z}
\hat{J}^{z}$ and the hf-mediated interaction from Eq.~(\ref{eq:H2}),
for which we obtain a perfect recovery of coherence in SE experiment.
However, the initial ``visibility loss''
corresponds to a tilt of the electron spin quantization axis from the
original $z$ direction by an angle proportional to $\delta$, and thus
the decrease of $\mean{\hat{S}^{z}}$ by a factor $\sim \! \delta^{2}$.
If we then rotate the coordinate system so that the new $z'$ direction
is along the new electron spin quantization axis, we will obtain the
terms of $\hat{S}^{z}{'}\hat{J}^{x}{'}$ type in $\hat{H}_{0}$ defined
above, thus arriving at the problem analogous to ESEEM due to
anisotropic hf interactions.

\begin{figure}
\includegraphics[width=0.99\linewidth]{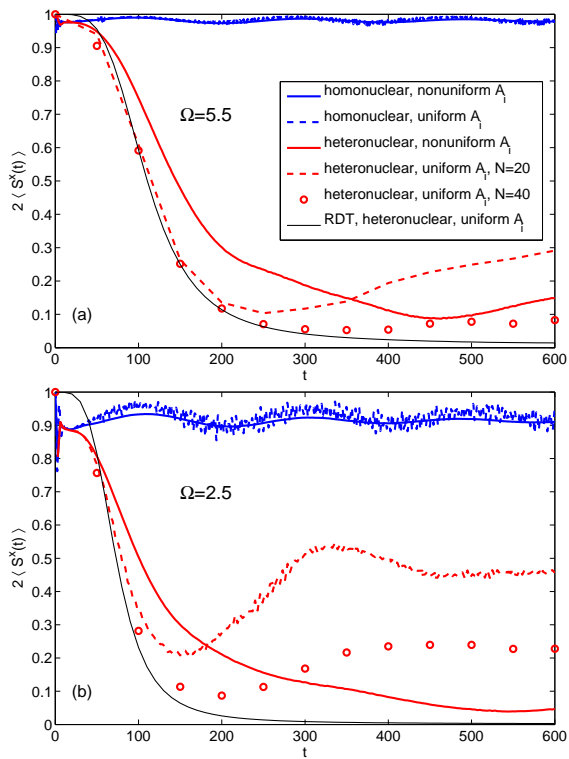}
\caption{(Color online) (a) Comparison of the SE signal calculated
with non-uniform hf couplings (red and blue solid lines) and with all
$A_{i} = A$  (dashed lines for $N=20$, circles for $N=40$) for
homonuclear and heteronuclear baths with $N\! = \! 20$ and at $\Omega
\! = \! 5.5$. The thin black solid line is the RDT result for uniform
couplings (which is $N$ independent). (b) The same for $\Omega \! = \!
2.5$.
} \label{fig:box_hetero}
\end{figure}

The ``box wavefunction'' approach was used previously to calculate
low-$B$ free evolution of $S^{z}$ component of the central
spin,\cite{Zhang_PRB06} and it was shown to agree with the exact
solution (for a system with inhomogeneous $A_{i}$) on the timescale of
$T^{*}_{2}\sim \sqrt{N}/\mathcal{A}$. Above we have seen that the box model
captures the main features
of the SE signal in a homonuclear bath on a much longer timescale,
which exceed not only $T^{*}_{2}$ but also $N/\mathcal{A}$.

Interestingly, this agreement between the ``box'' model and the exact
calculation breaks down at shorter timescale for the heteronuclear
bath. In Fig.~\ref{fig:box_hetero} we show the comparison between the
nonuniform $A_{i}$ calculations and the ``box'' results for homo- and
heteronuclear baths (at $\Omega\! = \! 2.5$, $5.5$ and with $N\! = \!
20 $, $40$ nuclei). The RDT results obtained with the uniform hf
couplings are also shown (this calculation gives the SE signal
independent of $N$). For $\Omega \! = \! 2.5$ there is an obvious
difference in long-time behavior between the ``box'' calculation and
the calculation with nonuniform $A_{i}$. The ``box'' calculation
exhibits a saturation of the SE signal at $t\! \sim \! 400$ (with the
signal staying close to this value for much longer times, not shown in
the Figure). However, the amplitude of this saturation decreases with
increasing $N$. While we have $2\mean{S^{x}}_{sat} \! \approx \! 0.46$
for $N\! = \! 20$, for $N\! = \! 40$ we obtain a value of about
$0.22$, suggesting that these results are affected by finite-$N$
effects. The reason for which the these effects are spoiling the
agreement between the exact and ``box'' calculation only in the
heteronuclear case remains to be further elucidated.

At higher $\Omega$, as in Fig.~\ref{fig:box_hetero}a, the finite-$N$
effect is weaker, and with $N \! = \! 40$ the RDT and the full-
Hamiltonian calculation agree very well for uniform $A_{i}$. Both of
these signals exhibit a somewhat faster decay than the exact result
with nonuniform couplings (which is well reproduced by the RDT
calculation with nonuniform $A_{i}$, as shown in
Fig.~\ref{fig:SE_hetero}b). This shows that at $N \! = \! 20$ the
choice of the specific set of $A_{i}$ couplings can have a visible
impact on the SE signal. This is discussed in more details in Appendix
\ref{app:choice}.

\section{Conclusions}
It this paper we have investigated the low magnetic field spin echo
(SE) signal of an electron spin interacting via the hyperfine (hf) coupling
with the nuclear bath. Intrabath dipolar interactions have been
neglected.
We have used three theoretical approaches: the exact numerical
solution for a relatively small nuclear bath, the analytical
theory based on resummation of ring diagrams,
\cite{Cywinski_PRL09,Cywinski_PRB09} and calculations in which
uniform hf coupling with all the nuclei was assumed.

The exact numerical calculation strongly suggests that the ring
diagram theory
(RDT) describes quantitatively the SE decay when the electron
Zeeman splitting $\Omega$ is much larger than the typical Overhauser
field $\mathcal{A}/\sqrt{N}$ (with $\mathcal{A}$ being the
total hyperfine interaction and $N$ being the number of nuclei).
The timescale $\tau_{R}$ on which RDT describes quantitatively the
SE decay well has been shown to visibly exceed $N/\mathcal{A}$.
A qualitative difference between the SE decay due to interaction with
the homonuclear and the heteronuclear bath (containing nuclei with
distinct Zeeman splittings) is still clearly visible in the numerical
calculation for $\Omega\! \approx \! \mathcal{A}/\sqrt{N}$
(corresponding to $\sim\! 10$ mT in large GaAs dots). Generally, we
have found that qualitative and even semi-quantitative predictions of
the RDT are robust down to these fields. The saturation of the spin
echo decay time at low magnetic fields had been predicted before using
the RDT,\cite{Cywinski_PRL09,Cywinski_PRB09} and our new results lead
us to expect that this behavior will be robust down to $B$ fields
corresponding to typical Overhauser field, i.e.~we predict the SE
decay to be practically  $B$-independent between $10$ and $100$ mT in
GaAs quantum dot with $\sim \! 10^{5}$ nuclei.

Also, we have found that the  SE signal in a homonuclear bath
oscillates with the Zeeman  frequency of the single present  nuclear
species. This  effect is related to the well-known ESEEM phenomenon,
but to our knowledge its presence has not been discussed for the  spin
echo signal in case of the isotropic hf coupling. This feature might
be observed in spin echo experiments on spin qubits in quantum dots
based on materials having a single nuclear species,
e.g.~silicon,\cite{Shaji_NP08,Nordberg_MOS_PRB09} carbon nanotubes
\cite{Churchill_PRL09} or graphene.\cite{Fischer_PRB09}
In a heteronuclear bath (e.g.~in GaAs) we have found that at very low
magnetic fields (smaller than the typical Overhauser field) the spin
echo signal exhibits strong oscillations in which the Zeeman
frequencies of all the nuclear species are present.

Finally, we have shown that using  a simplified model of uniform
hf couplings (corresponding to using a box-like electron
wavefunction) we can recover certain qualitative features
of the SE signal at timescale exceeding both $T^{*}_{2}$
and $N/\mathcal{A}$. While the nearly perfect disentanglement
of electron spin and nuclear bath by the SE sequence was discussed previously at
high magnetic fields,\cite{Shenvi_scaling_PRB05,Yao_PRB06}
this result of our work suggests that the dynamics of
coherence recovery in the SE experiment is quite closely
related to classical dynamics of coupled electron and nuclear
spins also at much lower magnetic fields. The implications of
this observation, and also the analysis of accuracy of
analytical theories using the effective hf-mediated
interaction for modeling of the narrowed free induction
decay, are left for future research.

\section{Acknowledgements}
This work is supported by LPS-NSA.
Work at the Ames Laboratory was supported by the Department
of Energy --- Basic Energy Sciences under Contract
No. DE-AC02-07CH11358.
{\L}C also acknowledges support from the
Homing programme of the Foundation for Polish
Science supported by the EEA Financial Mechanism.

\appendix

\section{Dependence of the spin echo signal on the choice of hyperfine
couplings}  \label{app:choice}
In this work we choose the hf couplings $A_{i}$ randomly from a
uniform distribution. The realistic distribution $\rho(A)$ of the
$A_{i}$ couplings is related to the shape of the envelope wavefunction
$\Psi(\mathbf{r})$ of the electron:
\beq
\rho(A) = \frac{1}{\nu_{0}}  \int_{V} \delta\left(A - \mathcal{A} |
\Psi(\mathbf{r}) |^{2} \right) d^{3}r \,\, , \label{eq:rho}
\eeq
where $V$ is the total volume, $\nu_{0}$ is the volume of the unit
cell, and the wavefunction normalization is $\int_{V}
|\Psi(\mathbf{r})|^{2} d\mathbf{r}= \nu_{0}$. For the wavefunction
being a two-dimensional Gaussian we have $\rho(A) \! \sim \! A^{-1}
\Theta(A_{\text{max}}-A)$, where $\Theta(x)$ is the Heaviside step
function and $A_{\text{max}}$ is the largest hf coupling (for the
nucleus located at the center of the wavefunction). $\rho(A)$ for a
more realistic electron envelope is given in
Ref.~\onlinecite{Cywinski_PRB09}, and it behaves in qualitatively
similar way. The key point is that if we are only concerned about the
most strongly coupled nuclei, with $A_{i}$ not much smaller than
$A_{\text{max}}$ (which naturally dominate the decay at short times,
and possibly determine most of the decay at low $B$ fields), the
approximation of $\rho(A)$ by a constant is reasonable.

The long-time dynamics of the electron spin was predicted to be
influenced by the shape, specifically the tails, of the electron
wavefunction,\cite{Erlingsson_PRB04,Al_Hassanieh_PRL06,Chen_PRB07,Coish_PRB08,Cywinski_PRB09,Coish_PRB10} and thus the details of the
distribution $\rho(A)$ for small $A$.  These features of $\rho(A)$ are
impossible to capture with only $N=20$ spins. We are, however, not
currently concerned about this. The recent
experiments\cite{Bluhm_arXiv10} are showing that the low $B$ field SE decay occurs
on timescale $\leq \! N/\mathcal{A}$, on which according to the RDT
the wavefunction shape is relatively unimportant. Thus the question of
low-field accuracy of the RDT for an uniform distribution of $A_{i}$
is well motivated.

\begin{figure}
\includegraphics[width=0.99\linewidth]{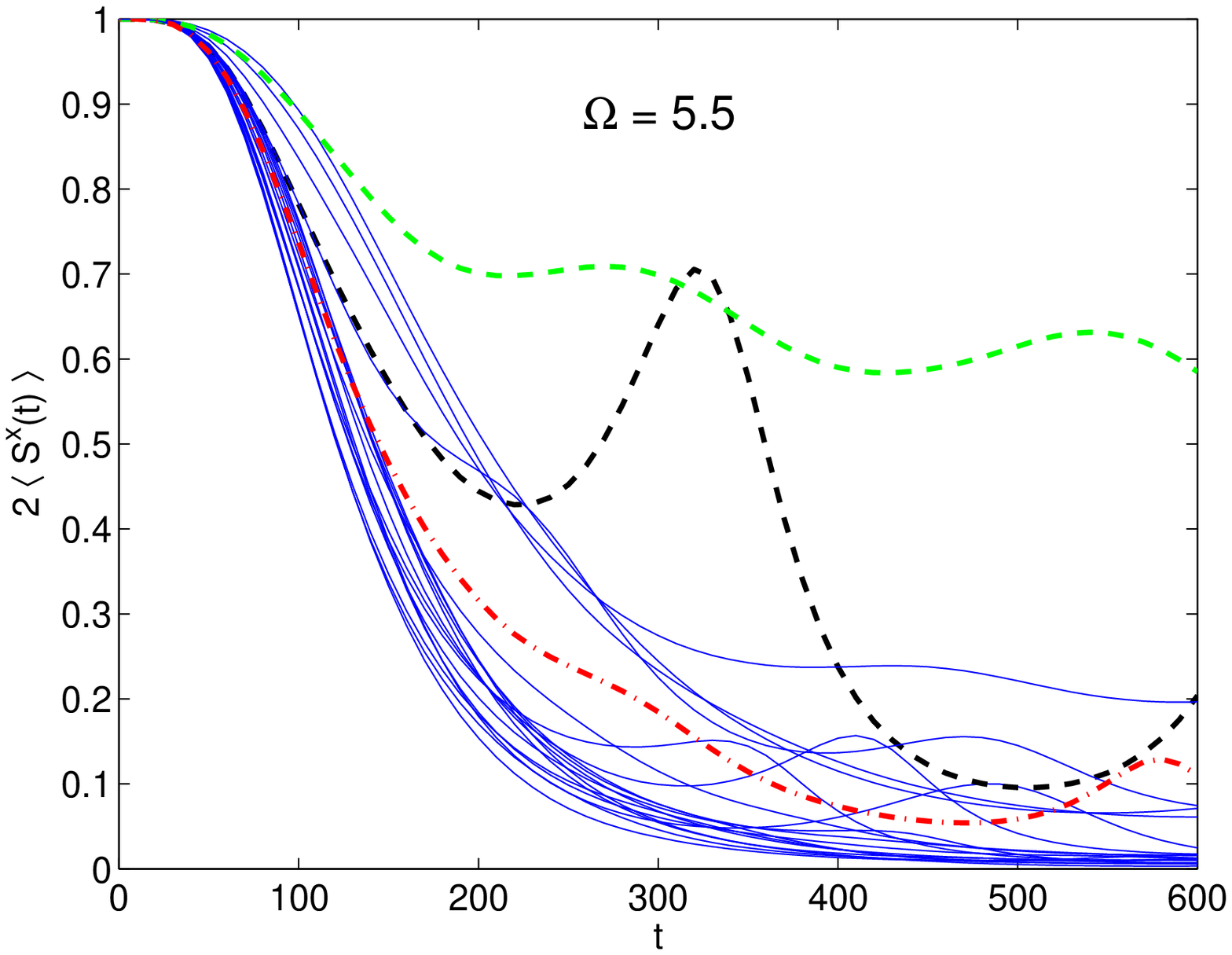}
\caption{(Color online) RDT calculation of the SE signal at $\Omega \!
= \! 5.5$ performed for $20$ different sets  of hf couplings $A_{i}$
(drawn at random from a uniform distribution, with $\sum_{i}A^{2}_{i}
\! =\! 1$ normalization). The signal for the set used in previous
figures is given by the red dot-dashed line. Two of the most
``atypical'' signals are drawn with dashed lines.
} \label{fig:choice}
\includegraphics[width=0.99\linewidth]{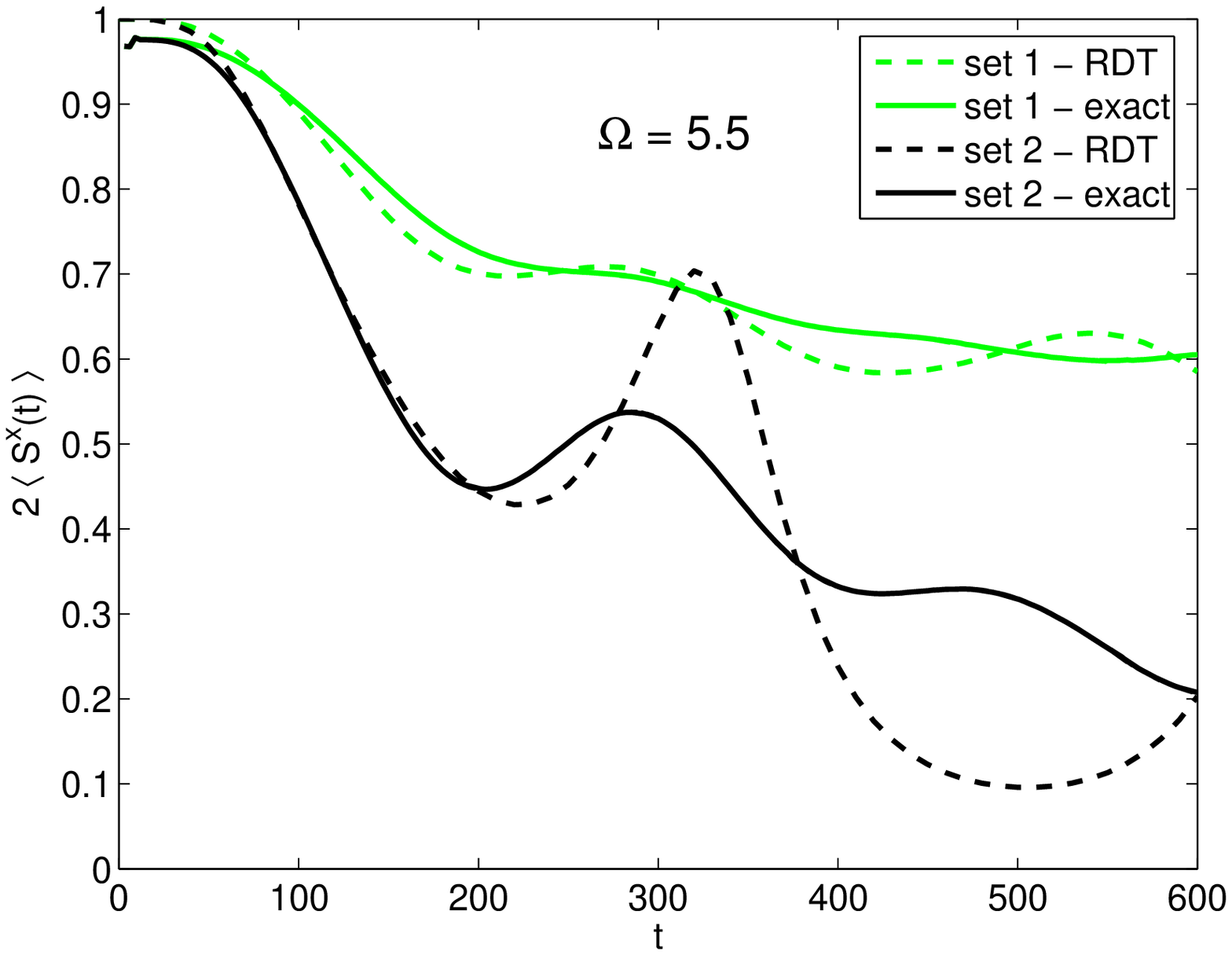}
\caption{(Color online) The comparison the the RDT calculation (dashed
lines) and the exact calculation (solid lines) for the two sets of hf
couplings which give atypical SE signals (shown in
Fig.~\ref{fig:choice} by dashed lines) } \label{fig:extreme}
\end{figure}

It is nevertheless prudent to check how our results depend on the
randomly chosen set of $A_{i}$. The reasonable expectation is that for
large $N$ the choice should not matter. This is \emph{not} the case
with $N \leq 10$, with which both the exact and RDT results are
showing very diverse behavior beyond the short time limit (i.e.~$t \!
< \! 100$ for $N\! = \! 10$). Unsurprisingly, for such small $N$ the
agreement between the two calculations is also present only for these
short times. For $N\! = \! 20$ the situation already becomes much
closer to our expectations. In Fig.~\ref{fig:choice} we present the
RDT results for $20$ random choices of the set of $A_{i}$ couplings at
$\Omega \! = \! 5.5$. One can see that the typical decay signal is
clearly visible. However, there are still sets of hf couplings which
give SE signals visibly differing from the typical behavior. The two
most extreme cases are drawn with dashed lines in
Fig.~\ref{fig:choice}. The set of $A_{i}$ used in previous
calculations corresponds to the dot-dashed line. This signal is quite
close to the typical one, although it does possess characteristic
features. These features closely correspond to the ones of the exact
signal shown before in Fig.~\ref{fig:SE_hetero}b, showing that at
$\Omega \! = \! 5.5$ the RDT calculation is very close to the exact
one for ``typical'' choice of $A_{i}$.

In Fig.~\ref{fig:extreme} we present the comparison between the RDT
and exact results for two sets of $A_{i}$ corresponding to most
atypical results from Fig.~\ref{fig:choice}. The values of hf
couplings used in these calculations are given in Table \ref{tab:Ai},
together with the set of $A_{i}$ used in most of the other figures in
the paper.
For the signal showing very weak decay (set 1) we find quite a good
agreement. For the signal exhibiting a sharp peak (set 2) the
agreement could be considered quantitative.

We have checked that the special behavior in case of set 2 is due to
presence of a large number of very small couplings in this set. In
fact, one can remove more than 10 smallest $A_{i}$ from this set
without visibly affecting the results, while the decay becomes
completely different (and much weaker) when one of the large
couplings, $A_{19}$, is removed.

In the case of set 1, the agreement is almost quantitative (only the
slow oscillation seems to be the artifact of RDT) As for the reason
for the exceptionally weak decay, it is possible that after removing
the $A_{i}$ couplings which are too small to visibly affect the signal,
the remaining couplings, especially the ones for different species of
nuclei (since hetero-nuclear interactions are crucial for SE decay)
are \emph{too uniform}. We note that as shown in Fig.~6, the decay for
$N=20$ with \emph{uniform} couplings saturates at about a half of
initial value of the signal. This saturation effect becomes weaker
when $N$ is increased.

\begin{table}
\begin{ruledtabular}
\begin{tabular}{cccc}
spin number $i$	&	$A_i$ (set 1) & $A_i$ (set 2) & $A_i$ (set 3)
\\
\hline
1  	& 	0.1236	&	  0.2681	&	0.2167	 \\
2	&  0.3984	&	0.3087	&    0.1746 \\
3	&	0.0376	&	0.1651		&	0.1033 \\
4	&	 0.1999	&	0.2836		&	0.2253 \\
5	&	0.0017	&	0.3709		&	0.2974 \\
6	&    0.1120	&	0.0088		& 	0.1769	\\
7	&    0.0007	&	0.0383		& 	0.2123 \\
8	&    0.0960	&	0.4457		& 	0.3171 \\
9	&    0.0723	&	0.2978		&	0.3479 \\
10	&    0.1360	&	0.1057		&	0.1659 \\
11	&    0.0887	&	0.1845		&	0.1655 \\
12	&    0.0704	&	0.0558		&	0.2818 \\
13	&   0.3039	&	0.1227		&	0.0928 \\
14	&   0.4572	&	0.1179		&	0.1386 \\
15	&	0.4767	&	0.1516		&	0.1520 \\
16	&    0.1122	&	0.0696		&	0.1225 \\
17  	&    0.2449	&	0.1591		&	0.0926 \\
18  &    0.1908	&	0.0556		& 	0.1416 \\
19  &   0.2658	&    0.4042	&    0.3951\\
20 &    0.1344	&	0.0431		&	0.3001\\
\end{tabular}
\end{ruledtabular}
\caption{Three sets of hf couplings $A_{i}$: sets $1$ and $2$
correspond, respectively,  to the most weakly decaying signal and to
the signal with prominent peak at $t \! \approx \! 350$ shown in
Fig.~\ref{fig:choice}, while set $3$ has been used in all the previous
figures. The values of $\omega_{i}$ nuclear Zeeman energies are:
$\omega_{i} \! =\! 0.02526$ for $i \! \leq \! 10$, $\omega_{i} \! =\!
0.0354$ for $i\! \in \![11,16]$ and $\omega_{i} \! =\! 0.045$ for $i\!
\in \![17,20]$.  } \label{tab:Ai}
\end{table}

It is also interesting to consider how the choice of the hf coupling
parameters affects the SE signal in the experimentally relevant limit
of very large $N$. We consider the SE signal $S(t)$, given by 
\begin{equation}
\label{eq:seslava}
S(2\tau) = {\rm Tr}[\hat S_x {\rm e}^{-i\hat H \tau}\hat\sigma_x
{\rm e}^{-i\hat H \tau} \hat S_x {\rm e}^{i\hat H \tau}\hat\sigma_x
{\rm e}^{i\hat H \tau}],
\end{equation}
which is just another way to write Eq.~(\ref{eq:Sx}). 
To access the limit of large $N$, we can use the Levy's lemma.
\cite{Levy,Popescu} It states that for a function $f(\vec x)$ defined on a
$(N-1)$-dimensional hypersphere $\vec x \in S^{N-1}$, and satisfying the 
1-Lipshitz condition, the value of the
function at a randomly chosen point is close to the
average value of this function $\langle f\rangle$ with very high probability:
\begin{equation}
{\rm Prob}[|f(\vec x)-\langle f\rangle|>\epsilon]\le 
\exp{(-CN\epsilon^2/L^2)}
\end{equation}
where $L={\rm sup}|\nabla f|$ is the Lipshitz constant. We use the
vector $(A_1, A_2,\dots A_k\dots, A_N)$ of the coupling constants,
with normalization $\sum_k A_k^2=1$, as a point $\vec x$ on a 
$N$-dimensional hypersphere, and the SE signal $S(t)$ 
as a function $f(\vec x)$. To 
evaluate $L$, we have to find the derivatives of 
$\partial S(t)/\partial A_k$, taking into account that the
only quantity dependent on $A_k$ in Eq.~(\ref{eq:seslava}) is the Hamiltonian $\hat H$. Thus, we need to substitute the equality
\begin{eqnarray}
\frac{\partial}{\partial A_k}{\rm e}^{-i\hat H \tau} &=& 
{\rm e}^{-i\hat H \tau} \int_0^\tau ds {\rm e}^{i\hat H s}
\frac{\partial\hat H}{\partial A_k} {\rm e}^{-i\hat H s} \\ \nonumber
&=& {\rm e}^{-i\hat H \tau} \int_0^\tau ds ({\mathbf S} {\mathbf J}_k)(s)
\end{eqnarray}
(and its Hermitian-conjugated version) into four places in 
Eq.~(\ref{eq:seslava}). In the Equation above  
$({\mathbf S} {\mathbf J}_k)(s) = {\rm e}^{i\hat H s} ({\mathbf S} {\mathbf J}_k) {\rm e}^{-i\hat H s}$.
Furthermore, we should take into account that
the trace of a matrix product ${\rm Tr}[\hat A^\dag \hat B]$ has all properties
of the scalar product, so that 
\begin{equation}
\left |{\rm Tr}[A^\dag B]\right |\le \sqrt{{\rm Tr}[\hat A^\dag \hat A] {\rm Tr}[\hat B^\dag \hat B]} = ||\hat A|| ||\hat B||,
\end{equation}
where $||A||=\sqrt{{\rm Tr}[A^\dag A]}$.
As a result, we obtain
\begin{equation}
\label{eq:cauchy}
\left |\frac{\partial S(t)}{\partial A_k} \right | \le 4 ||\hat S_x^2||
  \int_0^\tau ds ||({\mathbf S} {\mathbf J}_k)(s)|| = \sqrt{2}
  \int_0^\tau ds ||({\mathbf S} {\mathbf J}_k)(s)|| \,\, .
\end{equation}
Therefore, 
\begin{equation}
\left | \frac{\partial S(t)}{\partial A_k} \right | \le C_1 t
\end{equation}
with some constant $C_1$ independent of $N$, and for the Lipshitz constant we obtain
\begin{equation}
L \le C_1 t \sqrt{N}.
\end{equation}
Thus, for $t \! \sim \! 1$ (in our dimensionless units, where $\sum_k A_k^2=1$), the specific choice of the hf coupling parameters does not matter. 

However, it is obvious that our rigorous estimate Eq.~(\ref{eq:cauchy}), 
which is based on straightforward use of the Cauchy inequality, 
is too crude. Heuristically, we expect that the correlation between
the central spin and a single bath spin, which we crudely estimated 
from above as $O(1)$, is actually of order of $O(1/N)$. Then,
the range of times where the choice of the hf parameters does not 
matter, is extended to $t\sim N$.

\section{Spin echo in a model with uniform hf couplings}
\label{app:box}
We rewrite Eq.~(\ref{eq:Sx}) using the basis of $\ket{\pm,j,m}$ states
(with electron states $\ket{\pm x} \! = \! (\ket{+}\pm \ket{-})
/\sqrt{2}$) and with $\tau\! \equiv \! t/2$
\begin{eqnarray}
\mean{\hat{S}^{x}(t)} & = & \frac{1}{2^{N}} \sum_{j}\sum_{m=-j}^{j}
D_{j} \nonumber \\
& & \!\!\!\!\!\!\!\!\!\!\!\!\!\!\!\!\!\!\!\!\!\!\!\!\!\!\!\!\!\!
\bra{+x,j,m} e^{i\hat{H}t\tau}\hat{\sigma}_{x}  e^{i\hat{H}\tau}
\frac{ \hat{\sigma}_{x}}{2} e^{-i\hat{H}\tau} \hat{\sigma}_{x} e^{-
i\hat{H}\tau}  \ket{+x,j,m} \,\, ,  \label{eq:Sxbox}
\end{eqnarray}
where the sum over $j$ is from $0$ to $N/2$  ($1/2$ to $N/2$) for even
(odd) number of nuclei $N$, and the degeneracies of states with given
$j$ are given by\cite{Melikidze_PRB04}
\beq
D_{j} = \frac{N!}{(N/2-j)! (N/2+j)!} \frac{2j+1}{N/2+j+1} \,\, .
\eeq

As discussed in Sec.~\ref{sec:discussion}, the Hamiltonian couples
only pairs of $\ket{\pm,j,m}$ states:
\begin{eqnarray}
e^{-i\hat{H}\tau}\ket{+,j,m} = a_{jm}\ket{+,j,m} + b_{jm}\ket{-,j,m+1}
\,\, , \\
e^{-i\hat{H}\tau}\ket{-,j,m} = c_{jm}\ket{-,j,m} + d_{jm}\ket{+,j,m-1}
\,\, ,
\end{eqnarray}
and the coefficients $a_{jm}$ and $b_{jm}$ are given by
\begin{eqnarray}
a_{jm} & = & e^{-iE^{+}_{m} \tau} \left( \cos \frac{N^{+}_{jm}}{2}\tau
- i \frac{Z^{+}_{m}}{N^{+}_{jm}}  \sin \frac{N^{+}_{jm}}{2}\tau \right
) \,\, ,\\
b_{jm} & = & -i e^{-iE^{+}_{m} \tau}  \frac{X^{+}_{jm}}{N^{+}_{jm}}
\sin \frac{N^{+}_{jm}}{2}\tau \,\, ,
\end{eqnarray}
and $c_{jm}$ and $d_{jm}$ are given by analogous expressions with
superscripts $+$ replaced by $-$. $E^{\pm}_{m}$, $X^{\pm}$,$
Z^{\pm}$ and $N^{\pm}$ are given by
\begin{eqnarray}
E^{\pm}_{m} & = & [ (2m \pm 1) \omega - A/2] /2 \,\, , \\
X^{\pm}_{jm} & = & A \sqrt{j(j+1)-m(m\pm 1) } \,\, ,\\
Z^{\pm}_{m} & = & \pm [ \Omega - \omega + A(m\pm 1/2) ] \,\, , \\
N^{\pm}_{jm} & = & \sqrt{ ( X^{\pm}_{jm})^2 +  ( Z^{\pm}_{m})^2 } \,\,
.
\end{eqnarray}

Plugging these into Eq.~(\ref{eq:Sxbox}) and into an analogous
expression for  $\mean{\hat{S}^{y}(t)}$ (in which the middle
$\hat{\sigma}_{x}$ operator is replaced by $\hat{\sigma}_{y}$) we
arrive at the formula for the decoherence function
\begin{eqnarray}
W^{SE}(t) & = & \sum_{j=0}^{N/2}\sum_{m=-j}^{j} \frac{D_{j}}{2^{N}}
\Big( |a_{jm}c_{jm}|^{2} + a_{jm}c^{*}_{jm-1}|d_{jm}|^2 +  \nonumber
\\
& & a_{jm+1}c^{*}_{jm}|b_{jm}|^2    \Big ) \,\, .
\end{eqnarray}

In the case of heteronuclear bath with $N_{J}$ nuclear species one has
to introduce $N_{j}$ sets of basis states $\ket{j_{\alpha},m_{\alpha}}
$ in Eq.~(\ref{eq:Sxbox}). The Hamiltonian is then coupling the whole
subspaces of fixed $j_{\alpha}$, and one has to consider Hamiltonian
matrices of dimension $2\prod_{\alpha}(2j_{\alpha}+1)$ and evaluate
the evolution operators numerically.

\end{document}